\documentclass[12pt]{article}

\usepackage[english]{babel} 
\usepackage{times}
\usepackage{latexsym} 
\usepackage{amsmath}
\usepackage{amssymb}
\usepackage{graphicx}

\newcommand{\be}{\begin{equation}}
\newcommand{\ee}{\end{equation}}
\newcommand{\ba}{\begin{eqnarray}}
\newcommand{\ea}{\end{eqnarray}}

\let\f\frac

\begin{document}

\title
{\Large{\bf IMPOSSIBILITY of UNLIMITED GRAVITATIONAL COLLAPSE}}
\date{}
\author{\normalsize S.S.\,Gershtein,\footnote{E-mail:
Semen.Gershtein@ihep.ru}\,  A.A.\,Logunov,\footnote{E-mail:
Anatoly.Logunov@ihep.ru}\; M.A.\,Mestvirishvili
\\[2mm]
\small{\it Institute for High Energy Physics, 142281, Protvino,
Russia}}
\maketitle 

\begin{abstract}
{It is shown that the gravitational field, as a physical field
developing in the Minkowsky space, does not lead to unlimited
gravitational collapse of massive bodies and, hence, excludes 
a possibility of the formation of the ``black holes''.
}
\end{abstract}

In article [1] R. Tolman (see also [2]) had found, in the framework
of the general theory of the relativity (GTR), an exact, non-static
and
spherically-symmetric solution for the dust (zero pressure) according
to which   \textit{ the density achieves infinity}
during a finite proper time. The gravitational collapse of the
dust-like
sphere takes place. One usually believes that according the GRT a
massive star also collapses, when all sources of the thermonuclear
energy are exhausted, and a ``black holes'' is being formed. 

In this article we specially consider the question, in the framework
of  the relativistic  theory of gravitation (RTG)~[3], and in quite a
general form. RTG proceeds from the idea that the gravitational field
is universal and is a physical tensor field developing in the
Minkowsky space, whose source is the conserved energy-momentum tensor
of all matter fields including the gravitational one.  

Such an approach leads to the notion of an effective Riemann
space-time due to the presence of the gravitational field, and the
non-zero graviton mass is mandatory. 

The Lagrangean of the theory is fixed by the gauge group and the
requirement  that the gravitational field possesses only spins 2 and
0. The
complete system of generally covariant equations  
[3--5], form-invariant under the Lorentz transformation, is
 obtained from the least action principle. 

In such an approach, contrary to the GTR, the conservation laws for
the energy-momentum and angular momentum of all matter fields,
including the gravitational one, strictly hold. 

The system of equations of the RTG can be casted into a convenient
form 
\be
R_{\mu\nu}=8\pi\Bigl(T_{\mu\nu}-\f{\,1\,}{2}g_{\mu\nu} T\Bigr)
+\f{m^2}{2}(g_{\mu\nu}-\gamma_{\mu\nu}),
\label{eq1}
\ee
\be
D_\mu\,\tilde g^{\mu \nu} = 0\,. 
\label{eq2}
\ee
Here $g_{\mu\nu}$ is the metric tensor of the effective Riemann
space, 
$\gamma_{\mu\nu}$ is the metric tensor of the Minkowsky space,  
$\tilde g^{\mu \nu}=\sqrt{-g}\,g^{\mu\nu}$, $D_\mu$ is the covariant
derivative in the Minkowsky space, 
$m=m_g c/\hbar$, $m_g$ is the graviton rest mass. 

The equation for the substance 
( under "substance" we
understand all kinds of the matter except the gravitational field)
follows from Eq. (\ref{eq1}) and (\ref{eq2})
\[
\nabla_\nu T^{\mu\nu}=0\,,
\]
where  $\nabla_\nu$ is the covariant derivative in the Riemann space. 

One can see this if to make use of the Bianchi identity in Eq.(1):
\[
\nabla_\nu \Bigl(R^{\mu\nu}-\f{\,1\,}{2}g^{\mu\nu} R\Bigr)\equiv 0\,,
\]
and , having obtained  the equality following from some
transformations,
\[
m^2\gamma_{\alpha\lambda}g^{\alpha\mu}
D_\sigma\tilde{g}^{\lambda\sigma}
=16\pi \sqrt{-g}\,\nabla_\nu T^{\mu\nu},
\]
to substitute in it Eq.(2).

The energy-momentum tensor for the substance 
has the form
\be
T_{\mu\nu} =(\rho + p)U_\mu U_\nu -pg_{\mu\nu},\quad
T = T_{\mu\nu} g^{\mu \nu} =\rho -3p,
\label{eq3}
\ee
$\rho$ is the density of the substance, $p$ is the pressure. 
\be
U_\mu =g_{\mu\nu} \f{dx^\nu}{ds}.
\label{eq4}
\ee
Generic interval of the effective Riemann space is 
\be
ds^2 =g_{\mu\nu} dx^\mu dx^\nu .
\label{eq5}
\ee
We assume that the functions  $g_{\mu\nu}(x)$ belong to the class
$C^r$  and $r$ is defined in every concrete case. 
For localized timelike events $dx^i =0$ and we have
\be
ds^2 =g_{00} dt^2 > 0\,,
\label{eq6}
\ee
thereof 
\be
g_{00} > 0\,.
\label{eq7}
\ee
For simultaneous events $dt=0$ and the interval is spacelike  
\be
ds^2 =g_{ik} dx^i dx^k <0\,.
\label{eq8}
\ee
Thereof the Sylvester conditions follow 
\be
g_{11} < 0, \quad
\begin{vmatrix}
g_{11} & g_{12}\\
g_{21} & g_{22}\\ 
\end{vmatrix}
> 0\,,\quad
\begin{vmatrix}
g_{11} & g_{12} & g_{13}\\
g_{21} & g_{22} & g_{23}\\ 
g_{31} & g_{32} & g_{33}\\
\end{vmatrix}
< 0\,,
\label{eq9}
\ee
\be
ds^2 =d\tau^2 - dl^2\,.
\label{eq10}
\ee
Here 
\be
d\tau =\f{g_{0\nu}dx^\nu}{\sqrt{g_{00}}},\quad
dl^2 =\Bigl(-g_{ik}+\f{g_{0i}g_{0k}}{g_{00}}\Bigr)dx^i dx^k.
\label{eq11}
\ee
Conditions for the metric coefficient are the same both in the RTG
and GTR.

Below we will show , in the framework of the RTG and taking as an
example a \textit{massive non-static spherically-symmetric body},
that after
exhausting all nuclear resources no unlimited gravitational collapse
of the body occurs. 

The generic interval for the non-static spherically-symmetric
effective Riemann space can be presented in the form
\be
ds^2=g_{00}dt^2+2g_{01}dtdr+g_{11}dr^2+g_{22}d\theta^2+g_{33}d\phi^2,
\label{eq12}
\ee
where the metric coefficients  $g_{00}$, $g_{01}$, $g_{11}$  and
$g_{22}$ are the functions
of the radial variable $r$ and time $t$, while $g_{33}$ also depends
on the
angle $\theta$. 

We introduce the following notations
\ba
&&\!\!\!\!\!\!\!\!\!\!g_{00}(r,t)=U(r,t);\;g_{01}(r,t)=-A(r,t);\;
g_{11}(r,t)=-\bigl[V(r,t)-A^2(r,t)/U(r,t)\bigr];\nonumber \\
&&\!\!\!\!\!\!\!\!\!\!g_{22}(r,t)=-W(r,t);\;g_{33}(r,t,\theta)=-W(r,t
)\sin^2\theta\,.
\label{eq13}
\ea
The distance squared in the three-dimensional space is
\be
dl^2=Vdr^2 +W(d\theta^2 +\sin^2\theta d\phi^2),
\label{eq14}
\ee
and $V$ and $W$ are regular functions.

Non zero components of the tensor 
$g^{\mu\nu}$ are
\ba
&&\!\!\!\!\!\!\!\!\!g^{00}(r,t)=\Bigl(\f{1}{U}\Bigr)\Bigl(1-\f{A^2}{U
V}\Bigr);
\;g^{01}(r,t)=-\f{A}{UV};\nonumber \\[-2mm]
\label{eq15} \\[-2mm]
&&\!\!\!\!\!\!\!\!\!g^{11}(r,t)=-\f{1}{V};\;
g^{22}(r,t)=-\f{1}{W};\;g^{33}(r,t,\theta)=-\f{1}{W\sin^2\theta}\,,
\nonumber
\ea
while the determinant of the metric tensor is
\be
g=\det g_{\mu\nu} = -UVW^2\sin^2\theta.
\label{eq16}
\ee

For the interval of the Minkowski space  $d\sigma^2$
\be
d\sigma^2 =dt^2-dr^2-r^2(d\theta^2+\sin^2\theta\,d\phi^2),
\label{eq17}
\ee
we find, making use  of  (\ref{eq15})
\be
\gamma_{\mu\nu}g^{\mu\nu}=\f{\,1\,}{U}
\Bigl(1-\f{A^2}{UV}\Bigr)+\f{\,1\,}{V}
+\f{2r^2}{W}.
\label{eq18}
\ee
From the Sylvester conditions (9) we have according to (13) the
inequality 
\[
UV > A^2\,.
\]
It follows thereof and from  (7) that 
\begin{equation}
V>0.
\label{(19)}
\end{equation}

Contracting Eq.(10) with $g^{\mu\nu}$  and taking into account (18)
we find a
general relationship
\be
R + 8\pi T
=2m^2 -\f{m^2}{2}
\biggl[\f{\,1\,}{U}\Bigl(1-\f{A^2}{UV}\Bigl)+\f{\,1\,}{V}
+\f{2r^2}{W}\biggr].
\label{eq20}
\ee
It follows from this exact equality that $U$ and $V$ both inside and
outside the body cannot vanish, while function $W(r,t)$ cannot
vanish, at $r \to 0$, faster than $r^2$, otherwise $R$ or/and $T$
would get infinite 
and this would make impossible  sewing 
the solution inside the ball surrounded by the surface where physical
quantities are singular with the soultion outside this ball.

It worth noticing that while according to the GTR equations
\textit{the
metric coefficient $U$ can vanish}, it is impossible in the RTG which
corresponds to the general physical requirement (7).

Following Refs [6, 7] we take functions   $g_{\mu\nu}$ 
and $g^{\alpha\beta}$
from the class $C^3$, i.e.
they have continuous derivatives of the first three orders. Then,
according to Eq.(1), it follows that functions 
$R$, $\rho$ and $p$ 
are bounded inside the body. 
Let us note that their continuousness is
fairly sufficient for their boundedness.

Thus, both scalar curavature $R$ and $T$, and,  hence, according to
(3), density $\rho$ and pressure $p$ cannot get infinite, i.e. the
\textit{unlimited gravitational growth of the body density $\rho$ is
impossible and,
consequently, the gravitational collapse is absent}.

This result does not depend  of the
initial conditions.

As quantities $R$ and $T$ are invariants (scalars) no choice of
coordinates can modify this statement. For 
a static spherically-symmetric source, according to Eq.(2)
\[
A=0,
\]
and relationship (20) takes the following form
\[
R + 8\pi T
=2m^2 -\f{m^2}{2}
\biggl[\f{\,1\,}{U}+\f{\,1\,}{V}
+\f{2r^2}{W}\biggr].
\]

The essential feature is that general relationship (20) directly
follows from  gravitational equations (1) which are based on the
treatment of the gravitational field as a physical field in the
Minkowski space.

The field approach leads unavoidably to introducing the
\textit{graviton rest mass} with help of which the effective Riemann
space appeared related
to the Minkowski space.

Due to the graviton rest mass a relation arises, in the RTG, of the
scalar curvature $R$ and the substance invariant $T$ as exact
equality 
(20). It is this relationship which leads to the
\textit{impossibility of
the unlimited compression of the substance which forbids the
formation of the " black holes"}.

        The authors express their deep gratitude to Prof. V. A.
        Petrov and Prof. N. E. Tyurin for valuable discussions.

\renewcommand{\refname}


{\normalsize\bf References}
\begin{thebibliography}{99}
\bibitem{1}
\textit{Tolman\,R.C.} // Proc. Nat. Acad. Sci. 1934. Vol.~20.
P.~169--176. \bibitem{2}
\textit{Oppenheimer\,J.R., Snyder\,H.} // Phys. Rev. 1939. Vol.~56.
P.~455--459.
\bibitem{3}
\textit{A.A.Logunov} and \textit{M.A. Mestvirishvili}. Relativistic
Theory of
Gravitation.(In Russian). Moscow. Nauka, 1989.
\bibitem{4}
\textit{A. A.} Logunov. The Theory of Gravity. Moscow. Nauka, 2000.
\bibitem{5}
\textit{A. A. Logunov.} Relativistic Theory of Gravitation.(In
Russian).
Moscow. Nauka, 2006.
\bibitem{6}
\textit{V. A. Fock}. The Theory of Space, Time and Gravitation.
Pergamon
Press, 1964.
\bibitem{7}
\textit{J.L. Synge}. Relativity: the General Theory. North-Holland,
1960.
\end{thebibliography}
\end{document}